# Strong phase coherence and vortex matter in a fractal system with proximity-induced superconductivity

Nanami Teramachi,[1] Iku Nakaaki,[1] Aoi Hashimoto,[1] Shuuichi Ooi,[2] Minoru Tachiki,[2] Shunichi Arisawa,[3] Yusuke Seto,[4] Takahiro Sakurai,[5] Hitoshi Ohta,[6] Jaroslav Valenta,[2] Naohito Tsujii,[2] Takao Mori,[2,7] and Takashi Uchino[1]

[1] *Department of Chemistry, Graduate School of Science, Kobe University, Nada, Kobe 657-8501, Japan.*
[2] *International Center for Materials Nanoarchitectonics (WPI-MANA), National Institute for Materials Science, Tsukuba, Ibaraki 305-0047, Japan.*
[3] *Research Center for Functional Materials, National Institute for Materials Science, Tsukuba, Ibaraki 305-0047, Japan.*
[4] *Department of Geosciences, Graduate School of Science, Osaka Metropolitan University, Sumiyoshi, Osaka 558-8585, JAPAN*
[5] *Center for Support to Research and Education Activities, Kobe University, Nada, Kobe 657-8501, Japan.*
[6] *Molecular Photoscience Research Center, Kobe University, Nada, Kobe 657-8501, Japan.*
[7] *Graduate School of Pure and Applied Sciences, University of Tsukuba, Tsukuba, Ibaraki 305-8577, Japan*

**Abstract**

We investigate vortex matter in a proximity-coupled fractal system with ~30 vol. % of MgB$_2$ using magneto-optical imaging, scanning superconducting quantum interface device microscopy, and pinning force analysis. We show that this proximity-coupled system act as a fully phase coherent superconductor with isotropic pinning irrespective of the low volume fraction of MgB$_2$. Our results demonstrate that in contrast to the case of conventional granular superconductors, the grain boundaries in the present sample carry high critical currents and have high vortex pinning efficiency, implying an excellent phase-locked capability of the proximity-coupled fractal system.

When a normal metal (N) and a superconductor (S) are put in contact, their electronic properties are modified by their mutual interaction [1], leading to the penetration of superconducting correlations into the N material [2-4]. The phenomenon, which is known as the superconducting proximity effect [5,6], is a principal feature of Josephson junction-based devices [7] and has received renewed attention owing to the possibility of creating and manipulating Majorana fermions in topological insulator-S hybrid systems [8,9]. Important insights in the proximity effect are intimately connected to the process of Andreev reflection [6,10]. In this process, an electron coming from N with energy $\epsilon$ (measured with respect to the Fermi energy) is reflected as a hole of energy $-\epsilon$, resulting in the transfer of a Cooper pair of zero energy into the S region. In diffusive SNS junctions, the Andreev-reflected electrons and holes maintain phase coherence over a distance $L_C = \sqrt{\hbar D_N/\epsilon}$, where $D_N$ is the diffusion constant in the N region [6,10]. Thus, the coherence length $L_C$ for the pair-amplitude wave function diverges near the Fermi level $\varepsilon = 0$ [11]. However, the actual supercurrent flow is limited by several factors such as external magnetic fields [6], interface transparency [12], and the geometry of the junctions [13,14], which complicates the understanding of vortex behaviors in the proximity-induced superconducting regions. Although the vortices in tunnel Josephson junctions are believed to be coreless [15], recent theoretical and experimental investigations have provided evidence that in the long diffusive SNS junctions with highly transparent interfaces [16-18] or in some grain boundaries [19], the magnetic flux coming out of the N region is quantized to form proximity vortices containing cores with suppressed gaps. Hence, it is interesting to investigate the nature of flux patterns and the related critical current distributions in these SNS systems. However, no detailed report has been published on such issues, nor on the role of pinning force on the structure and properties of proximity vortices.

In this work, we investigate the superconducting and flux pinning properties of proximity-coupled superconductors by using an Mg/MgO/MgB$_2$ fractal nanocomposite with ~30 vol. % MgB$_2$, which can be viewed as a fractal assembly of long diffusive SNS junctions near or below the percolation threshold. Recently, we [20] have shown that such a fractal system behaves as a bulk type-II like superconductor from electrical, magnetic and muon spin rotation measurements, irrespective of the low volume fraction of superconducting (MgB$_2$) nanograins. Thus, this fractal nanocomposite provides an ideal experimental platform to explore the nature of proximity vortices and their pinning behavior. We here explore the flux patterns by means of magneto-optical (MO) imaging and scanning superconducting quantum interference device (SQUID) microscope methods. We also analyze the field and temperature dependence of the pinning force on the basis of a scaling law for flux pinning.

The bulk forms of the Mg/MgO/MgB$_2$ fractal nanocomposites with zero resistivity at 36.1 K [Fig. 1(a)] and perfect diamagnetism at 35.0 K [Fig. 1(b)] were synthesized by the solid phase reaction of Mg and B$_2$O$_3$ powders under Ar atmosphere at 700 ˚C, followed by a subsequent spark plasma sintering (SPS) procedure, basically according to the procedure reported in [20] (for details, see Supplemental Material [21]). The fractal structure, which is expected to be created during melting of B$_2$O$_3$ and the simultaneous redox reaction between B$_2$O$_3$ and Mg, was confirmed using a field-emission scanning electron microscope (FESEM)/energy-dispersive X-ray spectroscopy (EDX) imaging and a scanning transmission electron microscope (STEM)/EDX imaging. From the FESEM/EDX and STEM/EDX images with different magnifications [Fig. 1(c)], we see self-similar fractal structures consisting of highly complicated MgO-rich domains surrounded by barely interconnected MgB$_2$-rich regions irrespective of magnification. The fractal dimension of boron estimated from a box counting method is ~1.8 (Fig. S1 in the Supplemental Material [21]). From the X-ray diffraction pattern and the Rietveld pattern fitting, the composition was found to be MgO 75.8 wt% and MgB$_2$ 23.7 wt%, and Mg 0.5 wt% [Fig. 1(d)], corresponding to the approximate volume fractions of MgO 69 %, MgB$_2$ 30 % and Mg 1 %. It should be noted that the interface between the MgO- and MgB$_2$-rich regions are atomically clean and free of defects and voids, as seen in a high-resolution transmission electron microscope (HR-TEM) image [see the inset of Fig. 1(d)]. This clean interface will be essential to induce a robust and long-range proximity effect. If we apply the Ginzburg-Landau (GL) theory to this system and use the values of $H_{c1}(0) = 580$ Oe and $H_{c2}(0) = 110$ $k$Oe (Fig. S2 in the Supplemental Material [21]), the penetration depth λ and the coherence length ξ are estimated to be 89 nm and 5.5 nm, respectively. The resulting large GL parameter $\kappa = \lambda/\xi \sim 16 \gg 1$ is in consistent with the type-II like superconducting behaviors. We also measured the Hall coefficients ($R_H$) to identify the type of charge carrier and its density in the normal state (Fig. S3(a) in the Supplemental Material [21]). We found that $R_H$ is positive and increases with decreasing temperature, followed by a sudden decrease at temperatures below 30 K, as in the case of pure MgB$_2$ polycrystalline bulks [32]. This implies that the major carrier is the hole derived from MgB$_2$ nanograins although long-range electron transfer between oxygen vacancy defects in the MgO-rich regions [33] may play a role of normal conductors in the whole conduction process. We should also note that at 100 K, the net hole density $n_p$ is estimated to be $1.3 \times 10^{22}$ cm$^{-3}$ (Fig. S3(b) in the Supplemental Material [21]), which is an order smaller than that of pure MgB$_2$ ($n_p \sim 2 \times 10^{23}$ cm$^{-3}$ [32]).

Since we have confirmed that the present nanocomposite behaves as a bulk Type-II

like superconductor, we perform a series of zero-field-cooled (ZFC) MO measurements on the plate sample with dimensions of 5×5×1 mm$^3$ to study the flux penetration behavior. In the MO images, the penetration of the flux is imaged as bright areas, whereas the flux-free area stays dark. We found that at temperatures below 10 K, magnetic flux hardly penetrates into the sample on applying magnetic field $H$ up to 420 Oe except along the diagonal line (Movie S1 in the Supplemental Material [21]), which is due to an accidental physical crack created during surface polishing. When the temperature is increased to 30 K, magnetic flux penetration from the four edges can be recognized in the present experimental conditions (Fig. 2(a) and Movie S2 in the Supplemental Material [21]). A notable feature in this case is a rather smooth flux front [see the upper panel in Fig. 2(a)], unlike the case of ordinary polycrystalline and granular samples in which flux penetrates along the grain boundary or weak link admits vortices well before the bulk of the superconductor [30,34]. Figure 2(b) shows the flux-density $B$ profiles at 30 K obtained during the ramp-up stage of $H$. The observed flux penetration behavior is in harmony with the Bean critical state model [35] as the flux gradient is almost constant irrespective of $H$ especially for $H > 280$ Oe, yielding the local critical density $J_c$ of $1.0\times10^4$ A/cm$^2$ from the flux gradient ($\mu_0 J_c = dB/dx$). This value agrees well with that obtained from the magnetization measurements at 30 K, as will be shown later. Smooth flux fronts are also recognized in the MO images during the ramp-up and ramp-down stages of $H$ in the temperature range up to 35 K (Movies S3 and S4 in the Supplemental Material [21]). However, when the temperature is set to 36 K or above, the flux density almost corresponds to the applied field over the whole surface (Fig. 2(c) and Movies S5 and S6 in the Supplemental Material [21]). This indicates that at 36 K, the proximity induced $J_c$ becomes virtually zero.

We then performed the MO observations in the central area (135×135 $\mu m^2$) of the sample, where the effect of sample edge is neglected, after field cooling (FC) in 100 Oe to 30 K. Figure 3(a) demonstrates the resulting MO image, showing a rather homogeneous magnetic flux distribution. Note, however, that such a homogeneous flux distribution is almost retained even after switching off the field [Fig. 3(b)], demonstrating a nearly even spatial density of pinned vortices in the remanent state. Since the perpendicular flux distribution $B_z$ obtained from a thick-plate sample is generated mainly by the sum of a two-dimensional vortex current flow in the near-surface region [31], a numerical inversion of Biot-Savarts law can provide a map of the semi-quantitative current distribution in the sample surface [30] (for details, see Supplemental Material [21]). Figure 3(c) shows a mapping of the corresponding current vectors calculated by using the MO image shown in Fig. 3(b). One sees that the sum of the vortex currents flows continuously and windingly through the entire surface, suggesting not only the randomly pinned vortices, as observed in the flux-line vortex glass phase in a type-II superconductor, but also the strong intergrain phase coupling.

In a type-II superconductor, magnetic flux is quantized in units of flux quantum $\Phi_0$. It is hence interesting to investigate whether the magnetic flux trapped in the present nanocomposite is quantized or not. In order to clarify the issue, we observed scanning SQUID microscopy (SSM) images of the sample field-cooled to 3 K in 5 mOe, as shown in Fig. 4(a). The resulting SSM image reveals random distributions of bright dots with high magnetic flux density. Also, we found that each bright dot has a magnetic flux corresponding to one flux quantum $\Phi_0$, which was confirmed by integrating the observed $B_z$ values around the dot. Similar to the case of the MO images mentioned earlier, we can evaluate the local current distribution from the SSM image using a numerical inversion scheme of Biot-Savart law although the resulting current distribution will be expanded due to the extent of a stray field emanating from a vortex at the sample surface [36,37]. Figure 4(b) show the current distribution calculated from the MO image shown in Fig. 4(a). One can recognize typical features of a vortex structure, i.e., a central current-free region and its surrounding circulating current [see also the insets of Fig. 4(b)]. Hence, each bright dot shown in Fig. 4(a) will represent a single quantized vortex possibly with a core. After measuring the SSM image shown in Fig. 4(a), we then field-warmed the sample to 36 K, keeping the applied field at 5 mOe. The resulting temperature dependent SSM images are given in Fig. 4(c),(d). Although the border of the bright dots becomes blurred with increasing temperature, the overall dot pattern is almost preserved at temperatures up to 34 K [Fig. 4(c)]. When the temperature reaches 36 K, however, the dot signature is erased and is replaced with a nearly homogeneous pattern [Fig. 4(d)], implying complete suppression of the proximity induced $J_c$. This observation is fully consistent with the temperature dependent change in the MO images shown in Fig. 2. Hence, the bright dots shown in Fig. 4(a) can be regarded as proximity-induced quantized vortices that are pinned randomly in the N regions, supporting the vortex glass model.

To get further insight into the origin of the pinning, we performed the pinning force analysis using the critical current density $J_c$ derived from the height of the magnetization loop $\Delta M$ [inset of Fig. 5(a)] on the basis of the Bean model [35] (for details see the Supplemental Material [21]). The $J_c$ values in zero magnetic field are $\sim 1 \times 10^5$ and $\sim 3 \times 10^4$ A/cm$^2$ at 2 and 30 K, respectively [Fig. 5(a)]; the latter value is in reasonable agreement with that obtained from the MO images mentioned previously. We also found that similar to pure MgB$_2$ samples, the field dependence of $J_c$ is well fitted with the following equation:

$$J_c \propto \frac{[1 - H/H_{\text{irr}}]^2}{\sqrt{HH_{\text{irr}}}}, \quad (1)$$

where $H_{\text{irr}}$ is a fitting constant and also represents the field at which $J_c \to 0$, i.e., an irreversibility field [34,38,39]. This equation was originally developed for isotropic superconductors in which a dominant pinning mechanism is grain boundary pinning [40], and, in such a case, the volume pinning force density $F_p = \mu_0 H J_c$ shows a power law, $F_p \propto h^{0.5}(1-h)^2$, where $h$ is the reduced field $h = H/H_{\text{irr}}$. To investigate whether this isotropic pinning mechanism operates in the present sample, we examined the relationship between the reduced pinning force $f = F_p/F_{p,\text{max}}$ and $h$ [Fig. 5(b)]. We found that the $f$ curves obtained for temperatures at 2 and 30 K almost overlap and are well fitted to the function $h^p(1-h)^q$ with the following fitted values of $p$ and $q$: $p = 0.52$ and $q = 2.18$ for 2 K, and $p = 0.53$ and $q = 1.95$ for 30 K. The observed scaling behavior with $p \approx 0.5$ and $q \approx 2$ along with a well-defined linear behavior of the Kramer plot [$J_c^{0.5} \cdot H^{0.25}$ vs $\mu_0 H$, see the inset of Fig. 5(b)] allows us to confirm that the present sample is isotropic in terms of the pinning force, and also that the grain boundary pinning is the main cause of pinning as in the case of pure polycrystalline MgB$_2$ samples where grain boundaries do not limit the current flow [34]. We suggest that slight spatial modulations of the proximity-induced order parameter in N regions act as if they were natural and random grain boundary pinning sites.

The above experimental results, along with those of the MO and SSM observations, are rather surprising because in the conventional $MgB_2$ composites, the transition temperature $T_c$ and diamagnetic shielding properties suddenly decrease as the volume fraction of $MgB_2$ decreases due to the formation of heterograin weak boundaries [41,42]. Indeed, the bulk composite prepared by SPS of a simple mixture of Mg and MgO and $MgB_2$ powders shows a lower zero-resistivity temperature and weaker magnetic shielding than the fractal nanocomposite (Fig. S4 in the Supplemental Material [21]), confirming an important role of the fractality in achieving the global phase coherence. One possible scenario for the origin of the unusual phase coherence realized in the present nanocomposite is the formation of a fractal and hierarchical network of Andreev bound states (ABSs), which are electronic analogues of Fabry-Perot optical interferometers [43]. In such a network of ABSs, a fully phase locked state is expected to be achieved owing to the collective interference of the electron and hole states induced in each ABS, as occurred in coupled phase oscillators in a network with a complex interaction topology [44,45]. We should also note that randomness or disorder does not necessarily destroy phase coherence but may enhance superconducting correlations under some circumstances [46,47] , especially in systems with fractal [48-50] or multifractal [51,52] (i.e. a mixture of monofractals) characteristics. Moreover, scale-free fractal networks can induce an enhanced positive feedback mechanism, leading to strong correlation, self-organization, and network robustness [53,54]. Hence, we suggest that the global superconducting phase coherence realized in the present system arises from the collective synchronization of respective ABSs occurring at different length scales.

In summary, we have shown from the MO observations and the pinning force analysis that the present fractal nanocomposite behaves as if it were an isotropic type II superconductor in terms of magnetic flux expulsion, penetration and pinning. Also, the SSM images have elucidated that proximity-induced quantized vortices are pinned randomly in the N regions. These results demonstrate that the grain boundaries carry high critical currents and have excellent vortex pinning efficiency, resulting in the establishment of the global phase coherence throughout the sample. Our results not only challenge our conventional understanding of granular superconductors whose transport critical density and pinning strength are limited by weak intergrain coupling, but also reveal a high phase coherence capability of fractally coupled proximity network.

This research was carried under the joint research program of Molecular Photoscience Research Center, Kobe University with proposal numbers R03028. A part of this work was conducted in Institute for Molecular Science, supported by Nanotechnology Platform Program <Molecule and Material Synthesis> (JPMXP09S19MS1063b and JPMXP09S20MS1025) of the Ministry of Education, Culture, Sports, Science and Technology (MEXT), Japan. Spark plasma sintering was performed at Fuji Electronic Ind. Co., LTD., Saitama, Japan. We are indebted to H. Murakami, Osaka University, for providing us a MO plate. T. M. thanks support from JST Mirai JPMJMI19A1. T. U acknowledges financial support from a Grant-in-Aid for Scientific Research (21H01622) from the Japan Society for the Promotion of Science (JSPS), the Mitsubishi Foundation, and the Iketani Science and Technology Foundation.

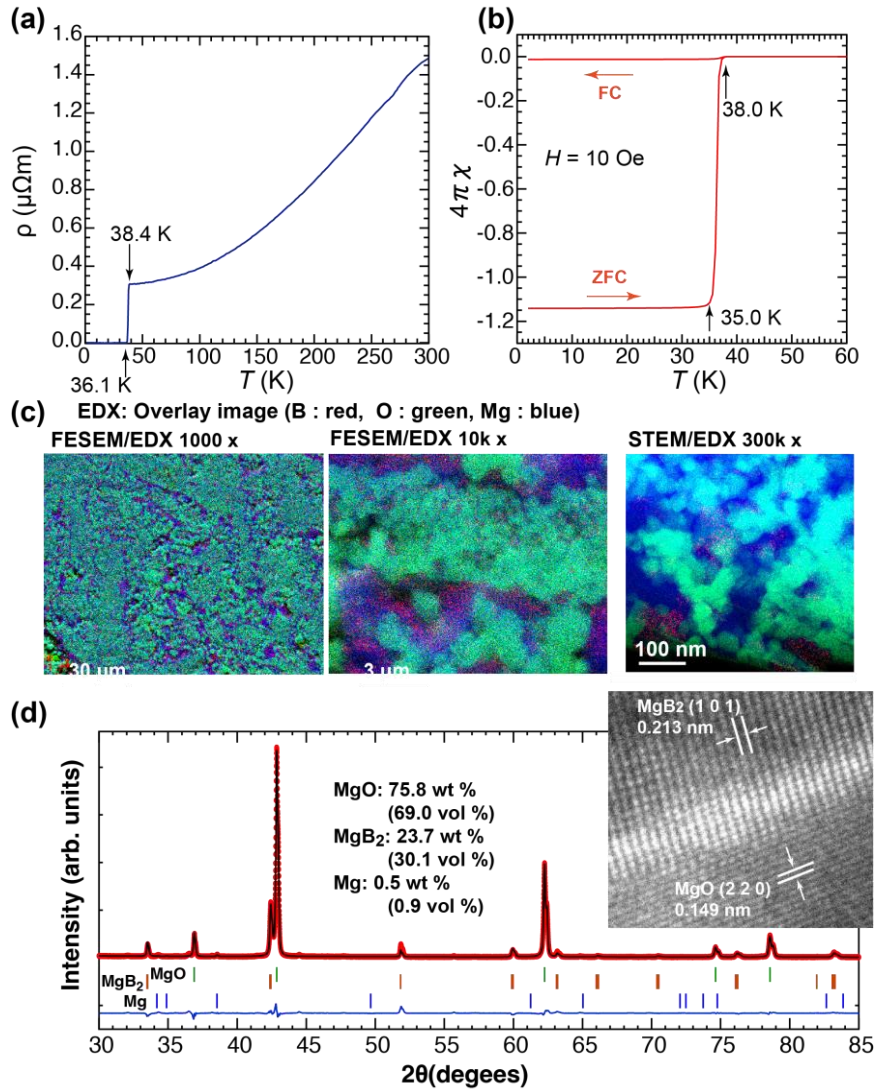

**FIG. 1.** (a) Temperature-dependent resistivity. (b) Zero-field-cooling (ZFC) and field-cooling (FC) magnetic susceptibility (4πχ) curves under applied field of 10 Oe. (c) FESEM/EDX and STEM/EDX overlay images in different magnifications; Red = B, Green = O, Blue = Mg. (d) XRD pattern and output from a quantitative Rietveld analysis. The lower curve is the difference between the observed data and calculated intensity at each step, plotted on the same scale. The inset shows a typical high resolution TEM (HR-TEM) image of the MgO/MgB$_2$ interface.

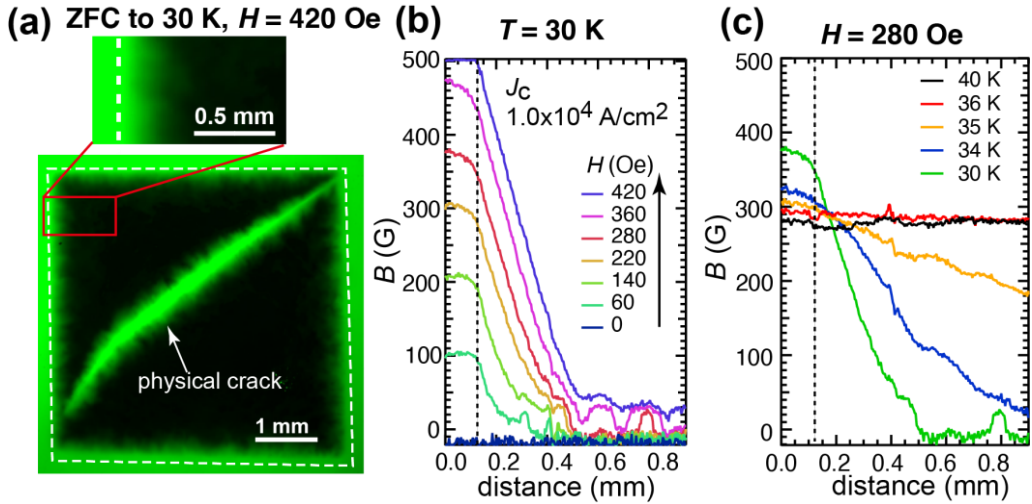

**FIG. 2.** MO observations after zero-field-cooling (ZFC) the sample. (a) MO image of the sample ZFC to 30 K and applying a magnetic field of 420 Oe. The edge of the sample is indicated by a dashed white line. The upper panel shows the enlarged image of the red box region. (b) Typical profiles of flux density at 30 K near the edge of the sample with increasing $H$ from 0 to 420 Oe. (c) Temperature dependence of the flux-density profiles observed in an applied field of 280 Oe. In (b) and (c), the dashed vertical line indicates the sample edge. The flux density near the sample edge exceeds the applied field due to the demagnetizing effect.

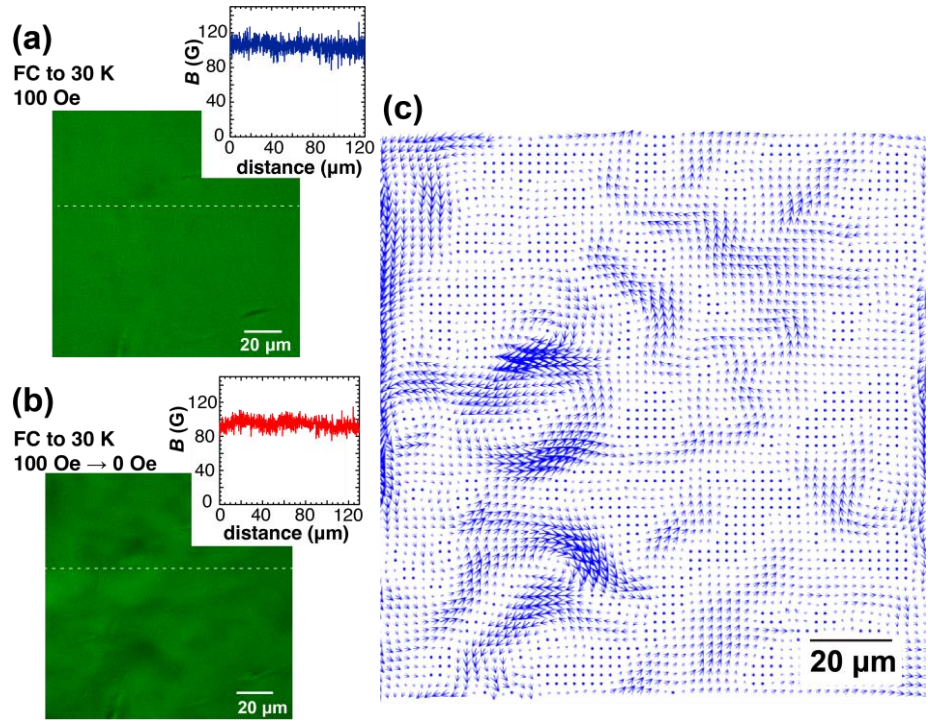

**FIG. 3.** MO images in the central area (135×135 μm$^2$) of the sample after (a) field-cooling (FC) to 30 K in an applied field of 100 Oe and (b) the subsequent switching off the field. The insets in (a),(b) show the flux-density profiles along the white dashed line in the respective MO images. (c) Current distribution pattern calculated by using the MO image shown in (b). The arrow length represent the magnitude of the absolute current density.

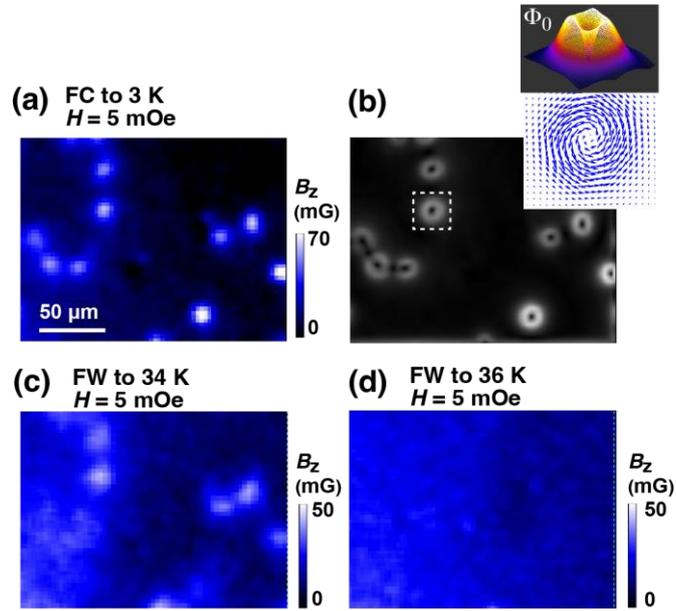

**FIG. 4.** (a) Scanning SQUID microscope (SSM) image taken after field cooling (FC) to 3 K in 5 mOe. (b) Supercurrents calculated from the SSM image shown in (a). The insets represent the local current distribution of the white box region in the forms of 3D false color image (upper inset) and current vectors (lower inset). The images (c),(d) show the SSM images taken after field-warming (FW) to (c) 34 K and (d) 35 K in 5 mOe.

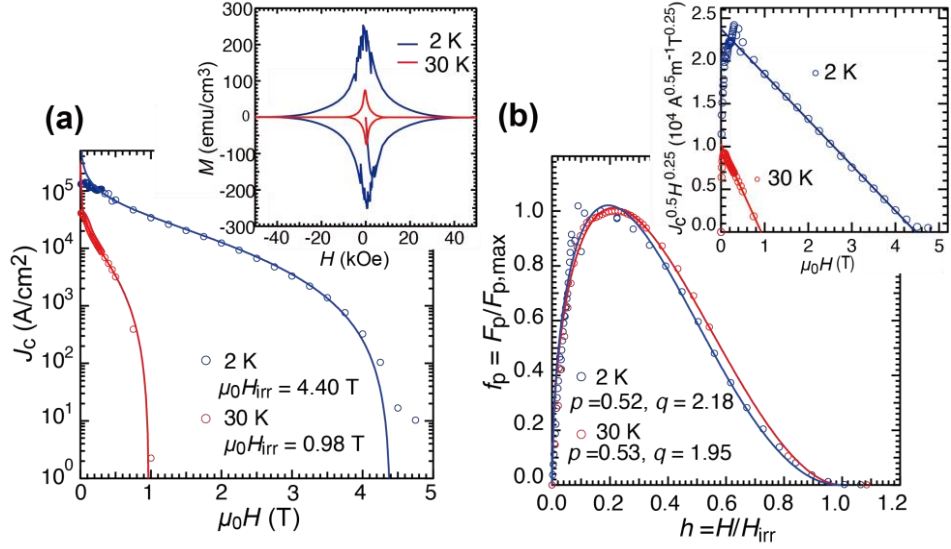

**FIG. 5.** (a) Critical current density $J_c$ at 2 and 30 K obtained from the corresponding $M(H)$ hysteresis loops given in the inset. The solid lines are fits of the data to equation (1), showing the fitted values of $\mu_0 H_{irr}$. (b) Normalized pinning force $f_p$ curves at 2 and 30 K. The lines are fits to the power law $f_p \propto h^p(1-h)^q$. The inset demonstrates Kramer plots, $J_c^{0.5} H^{0.25}$ vs $\mu_0 H$.